\documentclass[journal, letterpaper, 11pt]{IEEEtran}

\usepackage{url}
\usepackage{cite}
\usepackage{graphicx}
\usepackage{amssymb,amsmath,mathtools,amsthm,bm} %amscd,amsmath,verbatim
\usepackage[T1]{fontenc}
\usepackage{setspace}
\usepackage[dvips]{color}
\usepackage{multirow}
\usepackage{algorithm}
\usepackage{algpseudocode}
\usepackage{xcolor}
\usepackage{soul}
\usepackage[caption=false,labelfont=sf,textfont=sf]{subfig}
\newtheorem{Lem}{Lemma}
\newtheorem{Def}{Definition}

\newtheorem{Rem}{Remark}
\newcommand{\norm}[1]{\left\lVert#1\right\rVert}
\newcounter{MYtempeqncnt}

\begin{document}
\title{Quasi-Degradation Probability of Two-User NOMA over Rician Fading Channels}
\author{Kuang-Hao~Liu, \IEEEmembership{\small Member, IEEE}%
\thanks{Copyright (c) 2015 IEEE. Personal use of this material is permitted. However, permission to use this material for any other purposes must be obtained from the IEEE by sending a request to pubs-permissions@ieee.org.}
\thanks{Kuang-Hao~Liu (e-mail: {\tt khliu@mail.ncku.edu.tw}) is with the Department of Electrical Engineering,
National Cheng Kung University, Tainan, Taiwan 701.}
}
\date{}

\graphicspath{{./figure/}}

\maketitle

\begin{abstract}
Non-orthogonal multiple access (NOMA) has a great potential to offer a higher spectral efficiency of multi-user wireless networks than orthogonal multiples access (OMA). Previous work has established the condition, referred to quasi-degradation (QD) probability, under which NOMA has no performance loss compared to the capacity-achieving dirty paper coding for the two-user case. Existing results assume Rayleigh fading channels without line-of-sight (LOS). In many practical scenarios, the channel LOS component is critical to the link quality where the channel gain follows a Rician distribution instead of a Rayleigh distribution. In this work, we analyze the QD probability over multi-input and single-output (MISO) channels subject to Rician fading. The QD probability heavily depends on the angle between two user channels, which involves a matrix quadratic form in random vectors and a stochastic matrix. With the deterministic LOS component, the distribution of the matrix quadratic form is non-central that dramatically complicates the derivation of the QD probability. To remedy this difficulty, a series of approximations is proposed that yields a closed-form expression for the QD probability over MISO Rician channels. Numerical results are presented to assess the analysis accuracy and get insights into the optimality of NOMA over Rician fading channels.
\end{abstract}

\begin{keywords}
Gamma distribution, multi-input single-output (MISO), non-orthogonal multiple access (NOMA), quasi-degradation, Rician fading.
\end{keywords}

\section{Introduction}

Due to the scarcity of wireless spectrum, high spectral efficiency has been the primal design goal of contemporary wireless communication systems that impose multiple users to share the same spectrum. Traditionally, spectrum sharing is performed through orthogonal multiple access (OMA). Recently, non-orthogonal multiple access (NOMA) has received tremendous attentions from both academia and industry for its improved spectral efficiency over OMA~\cite{Saito2013}. The great potential of NOMA also stimulates the interest of the standardization body to include NOMA in the standard of the 5th generation (5G) mobile network~\cite{38812}.

While NOMA has been extensively studied in the literature, there remain numerous challenges to the successful application of NOMA in practice. One major concern is the implementation complexity. For power-domain NOMA, the superimposed user signals need to be decoded using successive interference cancellation (SIC) that increases receiver complexity. Also, the power allocated to different users is crucial to the achievable rate of NOMA. To determine the optimal power allocation, channel state information (CSI) is required at the transmitter side that is generally not accurate. Besides, the power allocation for sum rate maximization is a non-convex optimization problem that does not have closed-form solutions even for the two-user case~\cite{Sun2015}. Furthermore, there is no guarantee that the sum rate achieved by NOMA through optimal power allocation can approach the capacity region of multi-input and multi-output (MIMO) broadcast channels. For example, when different users' channels are nearly orthogonal or have similar channel gains, NOMA may not be preferred~\cite{Zhiguo2018}. 

Most existing work focuses on maximizing the sum rate for NOMA users by optimizing the power allocation and the decoding order for given user channel realizations. There is little discussion on whether the given user channels are attainable to achieving the capacity region. In~\cite{Xu2015}, the authors characterize the relationships among the capacity region of MIMO broadcast channels and rate regions achieved by NOMA and time-division multiple access (TDMA), which is a typical member of OMA family. For a given single-antenna user pair served by a single-antenna transmitter, the probability that NOMA outperforms TDMA in terms of sum rate and individual user rates is derived. In~\cite{Chen2016a}, the notion of quasi-degraded channels is introduced to evaluate the condition of users channels that achieve the capacity region of broadcast channels using NOMA. Since the exact capacity region of broadcast channels is not known, the sum rate of capacity achievable dirty paper coding (DPC) is considered as the benchmark~\cite{Weingarten2006}. For a pair of users, their channels are quasi-degraded if applying NOMA incurs no performance loss compared to that of using sophisticated DPC, which employs the encoding order same as the decoding order in NOMA. The notion of quasi-degradation (QD) is also exploited in~\cite{Chen2016} to address the user pairing and optimal precoding design for multi-user NOMA systems, where the precoders are designed for two users that are paired if their channels are quasi-degraded. Besides, the authors derive the QD probability when a multi-antenna base station (BS) serves two single-antenna users, i.e., a multi-input and single-output (MISO) setup. Recently, a new precoder design framework is propose in\cite{Zhu2021}, where QD is imposed as a constraint in the precoder design for intelligent reflecting surface (IRS) assisted NOMA. Different from the conventional multi-antenna systems that employ active antenna arrays, IRS is implemented using passive antennas that can smartly reflect the impinging electromagnetic waves to change the channel directions. With IRS, the user channels are tuned such that the QD can be satisfied. Thus the proposed design framework ensures that the obtained precoders achieve as good performance as DPC. All the existing work~\cite{Xu2015,Chen2016a,Chen2016,Zhu2021} assumes Rayleigh fading with no line-of-sight (LOS) in the propagation channel. For some scenarios suitable for NOMA, e.g., high-frequency millimeter wave (mmWave) communications~\cite{Anjinappa2018} and high-altitude unmanned aerial vehicles (UAVs) communications~\cite{Zeng2019}, the quality of a communication link is dominated by the LOS component in which case the fading statistics follow Rician distribution instead of Rayleigh distribution. The measurement results have revealed that the 28 GHz millimeter wave outdoor channels follow a Rician distribution with the Rician-factor ranging from 5-8 ~dB~\cite{Samim2016}.

In this paper, we extend the work~\cite{Chen2016} and intend to characterize the optimality of NOMA over Rician fading channels. Given two fixed users, we theoretically analyze the probability that the two user channels are quasi-degraded, namely, using NOMA to serve these two users can achieve the identical performance as non-linear DPC. In our work, a general model for the MISO Rician fading channel is considered, including a deterministic component that captures the azimuthal angle of the LOS signal and a non-deterministic component due to the randomness of NLOS. Unlike the Rayleigh fading case, the presence of LOS component raises numerous challenges to the characterization of QD probability. Firstly, the deterministic LOS component results in non-zero and distinct means for each element in the channel vector, i.e, the channel vectors are not isotropic. Consequently, the channel power of the MISO Rician channel has a non-central distribution whose statistical characteristics are difficult to be obtained. Secondly, the QD probability heavily depends on the squared cosine of the angle between two user channels, which can be represented in the matrix quadratic form~\cite{Chen2016} as given by $\mathbf{X}^H \mathbf{A} \mathbf{X}$ where $\mathbf{X}$ is a complex random vector and $\mathbf{A}$ is a symmetric matrix. When $\mathbf{X}$ is a vector of an isotropic channel, the distribution of the matrix quadratic form follows chi-squared distribution~\cite{Chen2016}, but this is not the case for the non-isotropic channel. Moreover, the matrix $\mathbf{A}$ in our work is stochastic but existing results on the distribution of the quadratic form are limited to the constant matrix $\mathbf{A}$. It is worth highlighting that there have been fruitful results on the distribution of the matrix quadratic form in random vector $\mathbf{X}$ when $\mathbf{X}$ is real~\cite{Mathai1992,Singull2012}. For complex random vectors, \cite{Ratnarajah2005} considers the quadratic form in a zero-mean complex random vector while the case of a non-zero mean complex random vector is studied in~\cite{Mohsenipour2012,Ducharme2016}. However, the matrix $\mathbf{A}$ considered in~\cite{Ratnarajah2005,Mohsenipour2012,Ducharme2016} is constant. To the best of our knowledge, the statistical properties of the matrix quadratic form with a stochastic matrix remain an open problem in the literature.

In view of the aforementioned difficulties, it is not plausible to find the exact QD probability over Rician fading channels. In this work, we propose an analytical framework that derives the QD probability based on some celebrated approximation techniques. Our contributions are three-fold.
\begin{itemize}
\item We show that the distribution of the quadratic form with random matrix $\mathbf{A}$ and complex vector $\mathbf{X}$ can be approximated to the gamma distribution. This is achieved with the aid of the second-order moment matching technique. To this end, we obtain the mean and the variance of the quadratic form with random matrix $\mathbf{A}$ and complex vector $\mathbf{X}$ by extending the existing results for constant $\mathbf{A}$ and real vector $\mathbf{X}$. 
\item We show that the channel power of the MISO Rician fading channel, which follows the non-central chi-square distribution, can be also well approximated to the gamma distribution. This approximation greatly facilitates the analysis of the QD probability that involves the ratio between the channel powers of two independent Rician channels. 
\item Using the approximated distributions aforementioned, we obtain the QD probability over MISO Rician fading channels. Numerical results are presented to validate the accuracy of the approximated QD probability and provide insights to the optimality of NOMA subject to Rician fading. It is shown that the strength of the LOS component relative to the NLOS component affects the QD probability in a different way, depending on the angular difference between the user channels. When the LOS paths have a larger angle difference, the QD probability increases with the LOS dominance but the trend reverses when the angle difference is smaller. This implies if two users are not close in the angular domain, NOMA over LOS fading channels is more likely to be capacity achieving. On the contrary, NOMA over NLOS fading channels is preferred if two users are close in the angular domain.
\end{itemize}

The remainder of this paper is organized as follows. Some useful distributions and important results on the distributions of the matrix quadratic form are first established in Sec.~\ref{sec: preliminary}. Sec.~\ref{sec:model} presents the MISO Rician fading channel model. Theoretical analysis for the QD probability over Rician fading channels is conducted in Sec.~\ref{sec:analysis}. Sec.~\ref{sec: gamma-quadratic-form} presents the gamma approximation for the stochastic quadratic form, which is used in Sec.~\ref{sec: approx-qd-prob} to obtain the approximated QD probability. Numerical results are presented and discussed in Sec.~\ref{sec:results}. Finally, concluding remarks are drawn in Sec.~\ref{sec:summary}. 

\emph{Notations}: In our notations, italic letters are used for scalars. Vectors and matrices are noted by bold-face letters. For a square matrix $\mathbf{A}$, $\mathbf{A}^{-1}$, $\mathrm{tr}(\mathbf{A})$, $\mathbf{A}^T$ and $\mathbf{A}^H$ denote its inverse, trace, transpose and conjugate transpose, respectively. $\mathbf{I}$ and $\mathbf{0}$ denote an identity matrix and an all-zero matrix, respectively. For a complex-valued vector $\mathbf{x}$, $\norm{\mathbf{x}}$ denotes its Euclidean norm. $\Gamma(\cdot)$ denoted the gamma function. The distribution of a circularly symmetric complex Gaussian random vector with mean $\boldsymbol{\mu}$ and covariance matrix $\boldsymbol{\Sigma}$ is denoted by $\mathcal{CN}(\boldsymbol{\mu},\boldsymbol{\Sigma})$, and `$\sim$' stands for `distributed as'. $\mathrm{E}[\cdot]$ and $\mathrm{V}[\cdot]$ denote the statistical expectation and variance, respectively. $\mathfrak{R}(\cdot)$ and $\mathfrak{I}(\cdot)$ denote the real and the imaginary part of a complex number. Finally, $\mathbb{C}^{m \times n}$ denotes the space of $m\times n$ complex-valued matrices.

\section{Preliminaries}\label{sec: preliminary}
In this section, we establish the distribution of the matrix quadratic form that serves as the core of the analysis for the QD probability in Sec.~\ref{sec:analysis}. Besides, some useful distributions and the second-order moment matching technique relevant to our work will be reviewed. 
%====================================
% Definition of Q_A(X)
%====================================
\begin{Def}\label{def: quadratic form}
%Another approximation employed in this work is the distribution of the matrix quadratic form. 
Given a multivariate random vector $\mathbf{X}=(x_1,\cdots,x_N)^T$ and a symmetric matrix $\mathbf{A}=(a_{ij})$, the quadratic form of $\mathbf{X}$ is defined as~\cite{Mathai1992}
\begin{equation}\label{eq: quadratic form}
Q_{\mathbf{A}}(\mathbf{X}) = \mathbf{X}^T \mathbf{A} \mathbf{X} = \sum_{i=1}^N \sum_{j=1}^N a_{ij} x_i x_j.
\end{equation}
\end{Def}

From (\ref{eq: quadratic form}), $Q_{\mathbf{A}}(\mathbf{X})$ is a scalar function of $\mathbf{X}$. When $\mathbf{X}$ is a real random vector and $\mathbf{A}$ is a constant matrix, the mean and variance of $Q_{\mathbf{A}}(\mathbf{X})$ are well known as given below.

%====================================
% Mean, Variance of Q_A(X)
%====================================
\begin{Lem}\label{lem: statistics of Q}
For a real random vector $\mathbf{X}$ with mean $\boldsymbol{\mu}$ and covariance matrix $\boldsymbol{\Sigma}$, the mean and variance of $Q_{\mathbf{A}}(\mathbf{X})$ is given by~\cite{Mathai1992}, 
\begin{align}
\mathrm{E}[Q_{\mathbf{A}}(\mathbf{X})] &= \mathrm{tr}(\mathbf{A} \boldsymbol{\Sigma}) + Q_{\mathbf{A}}(\boldsymbol{\mu}) \label{eq: mean Q_A(X)}\\
\mathrm{V}[Q_{\mathbf{A}}(\mathbf{X})] &=  2\mathrm{tr}( (\mathbf{A}\boldsymbol{\Sigma})^2 ) + 4 Q_{\mathbf{A} \boldsymbol{\Sigma} \mathbf{A}}(\boldsymbol{\mu^T}). \label{eq: var Q_A(X)}
\end{align}
\end{Lem}

The quadratic form $Q_{\mathbf{A}}(\mathbf{X})$ is useful for defining sums of squares.
%====================================
% Sum of squares
%====================================
\begin{Lem}\label{lem: Q to norm}
The quadratic form $Q_{\mathbf{A}}(\mathbf{X})$ reduces to the squared norm of $\mathbf{X}$ when $\mathbf{A}=\mathbf{I}_N$, i.e., $\mathbf{X}^H \mathbf{X}=x_1^2+\cdots+x_N^2=\norm{\mathbf{X}}^2$. If the entries of $\mathbf{X}$ are  independent but not necessarily identical (i.n.i.d.) Gaussian distributed with non-zero means and unit variance, then $\norm{\mathbf{X}}^2$ follows the non-central chi-squared ($\chi^2$) distributed with $2N$ degrees of freedom.  
\end{Lem} 

\begin{Rem}
The exact distribution of $Q_{\mathbf{A}}(\mathbf{X})$ is only known when $\mathbf{A}$ is deterministic. For a stochastic, matrix $\mathbf{A}$, approximations are required to obtain the statistics of $Q_{\mathbf{A}}(\mathbf{X})$
\end{Rem}

Although the distribution of the non-central $\chi^2$ distribution is well known, the PDF involves the modified Bessel function of the first kind that yields no closed-form expressions for the QD probability. For analytical tractability, we approximate the squared Gaussian random variables to be gamma distributed, which is defined as follows.

%====================================
% Gamma distribution
%====================================
\begin{Lem}\label{lem: gamma distri}
If $X$ is Gamma distributed with the shape parameter $k$ and the scale parameter $\theta$, the PDF is given by
\begin{equation*}
f_X(x) = \frac{1}{ \Gamma(k) \theta^k } x^{k-1} e^{-\frac{x}{\theta}},
\end{equation*}
where $\Gamma(\cdot)$ denotes the Gamma function. Moreover, $X \sim \Gamma(k,\theta)$ has the mean and variance given as
\begin{align*}
\mathrm{E}[X] &= k\theta, \\
\mathrm{V}[X] &= k \theta^2. 
\end{align*}
\end{Lem}
The approximation of the squared Gaussian random variable to gamma distribution is established based on the following lemma.

%====================================
% Squared gaussian to gamma  
%====================================
\begin{Lem}\label{lem: X^2 distribution}
If the random variable $X \sim \mathcal{N}(\mu, \sigma^2)$, then $X^2$ has the same first and second order statistics as $\Gamma(k, \theta)$ where
\begin{equation}\label{eq: X^2 gamma}
k=\frac{ (\sigma^2+\mu^2)^2 }{ 2\sigma^2 (1+2\mu^2)},~\text{and}~\theta=\frac{ 2\sigma^2 (1+2\mu^2)  }{ \sigma^2+\mu^2 }.
\end{equation}
\end{Lem}
\begin{IEEEproof}
See Appendix~\ref{app: squared gaussian approx}.
\end{IEEEproof}

We will need to work on the sum of i.n.i.d. gamma random variables. The exact distribution of the sum of i.n.i.d. gamma random variables can be obtained using numerical methods such as inverting the characteristic function or the saddle-point approximation~\cite{Murakami2015}. However, the distribution obtained from numerical computations does not permit a closed-form expression that is necessary to the derivation of the QD probability of interest. In this work, we resort to the second-moment matching technique to obtain the approximated distribution for the sum of i.n.i.d. gamma random variables~\cite{Seifi2016,Jr2011}.  

\begin{Lem}[Second-order moment matching]\label{lem: order matching}
Let $\{X_n\}_{n=1}^N$ be a set of $N$ i.n.i.d. gamma random variables where $X_n \sim \Gamma(k_n, \theta_n)$. Then the sum $Z=\sum_{n=1}^N X_n$ has the same first and second order statistics as a gamma random variable with the shape and scale parameters given as
\begin{equation}\label{eq: order matching parameters}
k = \frac{ (\sum_n k_n \theta_n )^2 }{ \sum_n k_n \theta_n^2 },~\text{and}~\theta=\frac{ \sum_n k_n \theta_n^2 }{ \sum_n k_n \theta_n }.
\end{equation}
\end{Lem}

%====================================
% Inverse Gamma 
%====================================
Our analysis for the QD probability also relies on the inverse gamma distribution\cite{Cook2008} given as follows.  
\begin{Lem}\label{lem: inverse Gamma dist}
If a random variable $Z \sim \Gamma(k, \theta)$, then $Z^{-1}$ follows the inverse gamma distribution with the PDF given as
\begin{equation}
f_{Z^{-1}}(z) = \frac{1}{\Gamma(k)\theta^k}z^{-k-1}e^{-\frac{1}{kz}}.
\end{equation}
The mean and the variance of $Z^{-1}$ are
\begin{equation}\label{eq: Z^-1 statistics}
\mathrm{E}[ Z^{-1} ] = \frac{1}{(k-1)\theta},~\mathrm{V}[ Z^{-1} ] = \frac{1}{ (k-1)^2(k-2) \theta^2 } 
\end{equation}
for $k>2$.
\end{Lem}

%====================================
% Ratio of Gammas
%====================================
Our analysis also involves the quotient of two independent gamma random variables.
\begin{Lem}\label{lem: ratio gamma}
Given $V \sim \Gamma(k_v, \theta_v)$ and $W \sim \Gamma(k_w, \theta_w)$, $\frac{V}{W}$ follows the Beta prime distribution or known as the inverted beta distribution with the PDF given by~\cite{Cordeiro2012}
\begin{align}\label{eq: Beta prime}
f_{V/W}(x) = \frac{ x^{\alpha_v-1} (1/\theta_v)^{\alpha_v} (1/\theta_w)^{\alpha_w} }{ (x/\theta_w+1/\theta_v)^{\alpha_v+\alpha_w} B(\alpha_v, \alpha_w) }
\end{align}  
where $B(x,y)=\int_0^1 t^{x-1} (1-t)^{y-1} \mathrm{d}t$ is the Beta function.
\end{Lem}

Finally, the random vector $\mathbf{X}$ in the quadratic form encountered in our analysis is complex. The following lemma adopted from~\cite{Ducharme2016} establishes the connection between a complex-valued random vector and its real-valued counterpart  

\begin{Lem}\label{def: complex rv}
A complex random vector $\mathbf{Z} \in \mathbb{C}^N$ is constructed from a pair $\mathsf{X}=(\mathbf{X}_1^T,\mathbf{X}_2^T)^T$ of real random vectors as 
\begin{equation}
\mathbf{Z}=\mathbf{X}_1+j \mathbf{X}_2
\end{equation}
where $\mathbf{X}_i \in \mathbb{R}^N$ for $i=1,2$. Equivalently, $\mathbf{Z}$ can be represented as a pair $\mathsf{Z}=(\mathbf{Z}^T,\mathbf{Z}^H)^T$. The connection between $\mathsf{Z}$ and $\mathsf{X}$ is 
\begin{equation}\label{eq: complex pair}
\mathsf{X}=\mathbf{M} \mathsf{Z},~\mathsf{Z}=\mathbf{M}^{-1}\mathsf{X}
\end{equation} 
where $\mathbf{M}$ is a $2N \times 2N$ matrix given by
\begin{equation}
\mathbf{M}=\frac{1}{2} \begin{bmatrix} \mathbf{I}_N & \mathbf{I}_N \\
-j \mathbf{I}_N & j \mathbf{I}_N
\end{bmatrix}.
\end{equation} 
Since $\mathbf{M}^{-1}=2\mathbf{M}$, we have $\mathsf{Z}=2\mathbf{M}^H \mathsf{X}$.
\end{Lem}

\section{Channel Model}~\label{sec:model}

Consider a downlink wireless network consisting of one BS and two user terminal. The BS has $N\geq 2$ antennas and each user  has a single antenna. The channel vector between a user and the BS follows the MISO Rician fading channel model given by 

%=== Fig. 1 ===
\begin{figure}[!t]
\centering
\includegraphics[width=0.9\columnwidth]{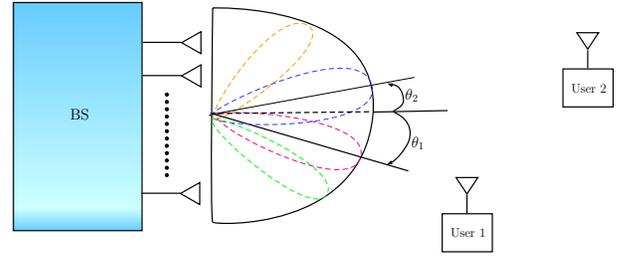}
\caption{Illustration of the two-user NOMA system. } \label{fig:model}
\end{figure}
\begin{equation}\label{eq: channel model}
\mathbf{g} = \sqrt{\beta}\left(\sqrt{\frac{1}{K+1}} \mathbf{h} + \sqrt{\frac{K}{K+1}} \mathbf{a}\right)
\end{equation}
where $\beta$ accounts for the large-scale fading due to pathloss, $\mathbf{h} \in \mathbb{C}^N \sim \mathcal{CN} (\mathbf{0}, \mathbf{I}_N )$ models the normalized NLOS component and $\mathbf{a} \in \mathbb{C}^N$ is a deterministic vector that captures the LOS component. The power ratio between the LOS component and NLOS component is determined by the Rician factor $K$. Assuming the $N$ BS antennas form a uniform linear array (ULA) with half-wavelength antenna spacing, the LOS component is modeled as $\mathbf{a} = [1, e^{-i \pi \sin(\theta)}, \cdots, e^{-i \pi (N-1) \sin(\theta)} ]^T$ where $\theta$ is the azimuth angle of the user. While the one-dimensional ULA is considered for its popularity, our work can be extended to other antenna array patterns such as uniform rectangular arrays (URAs) and uniform circular arrays (UCAs) with minor modifications. It is also worth stressing that our work generalizes the existing work~\cite{Chen2016} that considers Rayleigh fading channels with the Rician factor $K \rightarrow 0$.

In this work, we focus on two arbitrary users and characterize the QD probability, namely, the users channels permit the same performance using NOMA as that by DPC. We follow the definition of QD probability given in~\cite{Chen2016} for two users. Fig.~\ref{fig:model} illustrates the considered scenario. To perform NOMA, the BS transmits the combined signals of the two users $i$ and $j$ using superposition coding. Suppose a fixed decoding order $(i,j)$. User $i$ decodes the received signal by treating user $j$'s signal as noise. As to user $j$, it performs SIC by first decoding user $i$'s signal. Then user $j$ decodes its signal by subtracting user $i$'s signal from the received one. For detailed treatments on NOMA with SIC, we refer readers to \cite{Saito2013,Chen2016} and references therein.  

\section{Analysis of Quasi-degradation probability}\label{sec:analysis}

An efficient transmission scheme is important to the achievable capacity of the NOMA system. A common formulation for the optimal transmission scheme is to design the precoding vectors that minimize the transmission power subject to some quality-of-service (QoS) constraints~\cite{Chen2016,Chen2016a,Ding2019}. Specifically, for the two-user NOMA where the BS transmits the combined signals of the two users $i$ and $j$ using superposition coding, the precoding vectors are designed to minimize the transmission power subject to the target rate constraints $r_i$ and $r_j$. The same objective can be achieved by using DPC at the BS, where the encoding order $(i,j)$ is assumed to be fixed and identical to the decoding order of NOMA. For both NOMA and DPC, the design of the precoding vectors heavily depends on the users' channels. If the two users' channels denoted as $\mathbf{g}_i$ and $\mathbf{g}_j$, respectively, permit the same minimum transmission power of NOMA as that of DPC, their channels are quasi-degraded with respect to $r_i$ and $r_j$~\cite{Chen2016a}. It is shown in~\cite{Chen2016} that the condition of quasi-degraded channel can be expressed as
\begin{equation}
Q(\Theta) \leq \frac{ \norm{\mathbf{g}_i}^2 }{ \norm{\mathbf{g}_j}^2 },
\end{equation}
where 
\begin{equation}\label{eq: Q(Theta)}
Q(\Theta) = \frac{1+r_i}{\Theta}-\frac{r_i \Theta}{ (1+r_j(1-\Theta))^2 }
\end{equation}
and
\begin{equation}\label{eq: Theta}
\Theta= \frac{ \norm{\mathbf{g}_j^H \mathbf{g}_i}^2 }{ \norm{\mathbf{g}_i}^2 \norm{\mathbf{g}_j}^2 }.
\end{equation}

To illustrate the principle of quasi-degraded channel, a toy example is provided below. Two channel realizations $\mathbf{g}_i$ and $\mathbf{g}_j$ are generated according to the Rician channel model~(\ref{eq: channel model}) with the parameters following the simulation setup in Sec.~\ref{sec:results} with $K=5$ dB, $\theta_\Delta=10^\circ$ and $\beta_\Delta=100$.
\begin{equation*}
\mathbf{g}_i = \begin{bmatrix}
0.0483+0.0046i \\
0.0077-0.0377i\\
-0.0214-0.0217i\\
-0.0288+0.0161i
\end{bmatrix}, 
\mathbf{g}_j = \begin{bmatrix}
0.1846+0.0268i \\
0.066-0.0399i\\
-0.0515-0.0787i\\
0.1508-0.0428i
\end{bmatrix}
\end{equation*}
The given channel realizations have the gain power equal to $\norm{\mathbf{g}_i}^2=180.5$ and $\norm{\mathbf{g}_j}^2=13.5$, respectively. Besides, $\Theta=0.236$ and $Q(\Theta)=0.404$, which is less than $\tfrac{ \norm{\mathbf{g}_i}^2 }{ \norm{\mathbf{g}_j}^2 }=8.404$. Thus $\mathbf{g}_i$ and $\mathbf{g}_j$ satisfies the condition of quasi-degradation. It can be verified that the minimum power consumption by using NOMA to satisfy the rate constraints $r_i=r_j=1$ is identical to that using DPC as follows. According to~\cite{Chen2016}, we can obtain the precoders for minimizing the power consumption of NOMA, namely $\mathbf{w}_i$ and $\mathbf{w}_j$ for user $i$ and user $j$, respectively. Then the power consumption of NOMA is $P^{\text{DPC}}=\norm{\mathbf{w}_i}^2+\norm{\mathbf{w}_j}^2=0.08$. On the other hand, the optimal power consumption using DPC is \cite{Chen2016}
\begin{equation*}
P^{\text{DPC}}=\frac{ r_j }{ \norm{\mathbf{g}_j}^2 }+\frac{ r_i }{ \norm{ \mathbf{g}_i }^2 }\frac{1+r_j}{ 1+r_j \sin(\Theta) } = 0.08.
\end{equation*}
The above example demonstrates that the notion of quasi-degradation is useful to evaluate the optimality of NOMA in approaching the performance of non-linear DPC.

For convenience, denote the power ratio of the users' channels as 
\begin{equation*}
\frac{ \norm{\mathbf{g}_i}^2 }{ \norm{\mathbf{g}_j}^2 } \triangleq \Xi.
\end{equation*}
Then the broadcast channels $\mathbf{g}_i$ and $\mathbf{g}_j$ are quasi-degraded with probability~\cite{Chen2016}
\begin{align}\label{eq:P_QD_orig}
P_{QD}^{(i,j)} &=P[Q(\Theta) \leq \Xi] \nonumber \\
&= \mathrm{E}_\Theta[ \mathit{P}[ \Xi \geq Q(\Theta) \vert \Theta ] ] \nonumber \\
&= \int_0^1 \int_{Q(\vartheta)}^\infty f_{\mathbf{\Xi}}(\xi) \mathrm{d} \xi f_\Theta(\vartheta) \mathrm{d} \vartheta.
\end{align}
In the sequel, we derive the PDFs of $\Xi$ and $\Theta$ that are essential to the evaluation of $P_{QD}^{(i,j)}$. To begin with, we first analyze the distribution of the channel power for the considered Rician MISO channel (\ref{eq: channel model}). It is worth mentioning that when the channel vector is subject to different path-loss and LOS components as considered in this work, the distribution of channel power is non-isotropic. 

%====================================
% Distribution of |h|^2
%====================================
\subsection{Gamma Approximation for Channel Power}\label{sec: gamma approx channel power}\label{sec: channel power approx}
We first establish the approximated distribution of the channel power $\norm{\mathbf{g}}^2$ for the \ul{MISO} Rician channel in (\ref{eq: channel model}), which serves as the root of our work. Without loss of generality, the user index is dropped from the channel vector. Recall that each entry in $\mathbf{g}$ is a complex Gaussian random variable and thus the $n$th entry can be expressed as 
\begin{equation}
\mathbf{g}_n = X_n + i Y_n,~n=0, 1, \cdots, N-1.  
\end{equation}
Since $e^{-i x} = \cos(x)-i\sin(x)$, we have
\begin{align}
X_n &= \sqrt{\frac{\beta K}{K+1}} \cos( \varphi_n )  + \sqrt{ \frac{\beta }{K+1}} a_n  \nonumber \\
Y_n &= -\sqrt{\frac{\beta K}{K+1}} \sin( \varphi_n )  + \sqrt{ \frac{\beta }{K+1}} b_n
\end{align} 
where $\varphi_n = (n-1)\pi \sin(\theta)$, $a_n$ and $b_n$ are independent and follow the zero mean Gaussian distribution with variance $1/2$. Together with the fact that $\varphi_n$ is deterministic, both $X_n$ and $Y_n$ are Gaussian distributed as given by
\begin{align}
X_n &\sim \mathcal{N}\left( \sqrt{\frac{\beta K}{K+1}} \cos( \varphi_n ), \frac{ \beta }{2(K+1)} \right), \label{eq: Xn} \\
Y_n &\sim \mathcal{N}\left( -\sqrt{\frac{\beta K}{K+1}} \sin( \varphi_n ), \frac{\beta }{2(K+1)} \right) \label{eq: Yn}
\end{align}

With each entry in $\mathbf{g}$ characterized above, the channel power $\norm{\mathbf{g}}^2$ is the sum of squared complex Gaussian random variables with distinct means and the same variance. 
When $X_n$ and $Y_n$ have unit variances, $\norm{\mathbf{g}}^2$ follows the non-central $\chi^2$ distribution with $2N$ degrees of freedom according to Lemma~\ref{lem: Q to norm}. %While the PDF of the non-central $\chi^2$ distribution is well-known, it involves the modified Bessel function of the first kind that complicates the theoretical analysis of the QD probability. Instead of using the exact distribution, 
Unfortunately, the unit variance condition is valid only when the Rician factor $K=-1/2$, which is not realistic. Motivated by the gamma approximation for the non-isotropic fading channels in~\cite{Hosseini2016}, we propose to approximate $\norm{\mathbf{g}}^2$ as follows. Firstly, $X_n^2$ and $Y_n^2$ are approximated as two independent gamma random variables that are fully characterized by the first two moments with closed form expressions using Lemma~\ref{lem: X^2 distribution}. Then the channel power is approximately gamma distributed according to Lemma~\ref{lem: order matching}.

\begin{Rem}[Equivalence to the Rayleigh fading case]
When $K=0$, i.e., the Rician channel (\ref{eq: channel model}) reduces to the Rayleigh channel the gamma approximation leads to the same result as~\cite{Chen2016}, which derives the QD probability for the Rayleigh channel by considering the fact that $\norm{\mathbf{g}_i}^2$ is a chi-square random variable with $2N$ degrees of freedom. For the Rayleigh channel, $X_n$ and $Y_n$ are both zero mean Gaussian random variables. According to Lemma~\ref{lem: X^2 distribution}, $X_n^2$ and $Y_n^2$ have the same first and second order statistics, e.g., $\Gamma(\frac{1}{2},2)$ for $\beta=1$. Using the second-order moment matching in Lemma~\ref{lem: order matching}, $\norm{\mathbf{g}_i}^2$ as the sum of $2N$ gamma random variables can be approximated as a gamma random variable with the shape parameter $k=N$ and scale parameter $\theta=2$. Since $\Gamma(N,2)$ is also a chi-square random variable with $2N$ degrees of freedom, our analysis is equivalent to that in~\cite{Chen2016} for Rayleigh channels.
\end{Rem}

\begin{Rem}[Hassle of the exact distribution]
The exact distribution of $\norm{\mathbf{g}}^2$ can be obtained using the result in~\cite{Tavares2007}, which represents $\norm{\mathbf{g}}^2$ in the Eular form and gives the distribution of the amplitude and the phase components. In their results, the Fourier series representation is used to numerically evaluate an improper integral appeared in the density function of $\norm{\mathbf{g}}^2$. Hence, their results are not useful to our work.
\end{Rem}

%====================================
% Xi = |g_i|^2/||^2
%====================================
\subsection{Distribution of $\Xi$}
Given that the user's channel power is approximated to the gamma distribution, $\Xi$ as the ratio of $\norm{\mathbf{g}_i}^2$ and $\norm{\mathbf{g}_j}^2$ follows the Beta prime distribution according to Lemma~\ref{lem: ratio gamma}.

%====================================
% PDF of \Theta
%====================================
\subsection{Distribution of $\Theta$}
First we examine the structure of $\Theta$. From (\ref{eq: Theta}), the numerator of $\Theta$ is the squared norm of the inner product between user $i$'s and user $j$'s channel vectors. The channel vector contains the complex entries with non-central means, in which case the PDF of the numerator of $\Theta$ involves complicated expressions. On the other hand, the denominator is the product of the two users' channel power. As mentioned in Sec.~\ref{sec: channel power approx}, each channel power is non-central $\chi^2$ distributed. The PDF of the product of two independent non-central $\chi^2$ random variables is given in~\cite{Wells1962}, which involves the infinite sum of the modified Bessel function of the second kind. 

From the above discussion, the original form of $\Theta$ in (\ref{eq:P_QD_orig}) is analytically non-tractable. Alternatively, we resort to an equivalent form as given by
\begin{equation}\label{eq: revised u}
\Theta = \frac{ \mathbf{g}_i^H \mathbf{g}_j \mathbf{g}_j^H \mathbf{g}_i }{ \norm{\mathbf{g}_i}^2 \norm{\mathbf{g}_j}^2 } = \frac{ \mathbf{g}_i^H \boldsymbol{\Pi}_{\mathbf{g}_j} \mathbf{g}_i }{ \norm{\mathbf{g}_i}^2 }
\end{equation}
where $\boldsymbol{\Pi}_{\mathbf{g}_j} = \mathbf{g}_j(\mathbf{g}_j^H \mathbf{g}_j)^{-1} \mathbf{g}_j^H$. In~(\ref{eq: revised u}),  the denominator contains the channel power of one user only and thus is simpler than the original form. From Sec.~\ref{sec: channel power approx}, $\norm{\mathbf{g}_i}^2$ can be approximated as a gamma random variable. On the other hand, the numerator follows a matrix quadratic form of the random vector $\mathbf{g}_i$. Notice that $\boldsymbol{\Pi}_{\mathbf{g}_j}$ is a stochastic matrix whose exact distribution is not tractable. Motivated by the gamma approximation for $\norm{\mathbf{g}_i}^2$, we also approximate the numerator of $\Theta$ by gamma distribution. The approximated distributions greatly simplify the analysis yet allow for numerical evaluation of the QD probability with reasonable accuracy. % under some conditions. 
We note that when the user channel is subject to Rayleigh fading without the LOS component, the numerator of $\Theta$ is $\chi^2$ distributed with degree of freedom 2~\cite{Chen2016}.

\section{Gamma Approximation for the Stochastic Quadratic Form}\label{sec: gamma-quadratic-form}
The numerator of $\Theta$ can be expressed in the quadratic form as $Q_{\mathbf{A}}(\mathbf{X})$ where $\mathbf{A}=\boldsymbol{\Pi}_{\mathbf{g}_j} = \mathbf{g}_j (\mathbf{g}_j^H \mathbf{g}_j)^{-1} \mathbf{g}_j^H$ and $\mathbf{X}=\mathbf{g}_i$. Since $\mathbf{g}_i$ is a complex random vector and $\boldsymbol{\Pi}_{\mathbf{g}_j}$ is stochastic, existing results shown in Lemma~\ref{lem: statistics of Q} for the real vector $\mathbf{X}$ and constant matrix $\mathbf{A}$ can not be used directly. In this section, we derive the mean and the variance of the complex quadratic form $Q_{\mathbf{S}}( \mathbf{Z} )$, where $\mathbf{S}$ is a stochastic matrix and $\mathbf{Z}$ is a complex vector. Then the numerator of $\Theta$ in (\ref{eq: revised u}) is approximated as a gamma random variable with the shape and scale parameters determined by the mean and variance obtained in the following.

\subsection{Mean}\label{sec: mean of Q_Pi(g)} %$Q_{\boldsymbol{\Pi}_\mathbf{g}}(\mathbf{g})$
We first establish the mean of the complex quadratic form $Q_{\mathbf{A}}( \mathbf{Z} )$ when $\mathbf{A}$ is deterministic and $\mathbf{Z} \in \mathbb{C}^N$. 
%====================================
% Mean of Q_A(Z)
%====================================
\begin{Lem}\label{lem: mean Q_A(Z)}
Consider a complex random vector $\mathbf{Z}$ where the real and the imaginary parts have the same distribution. For a deterministic matrix $\mathbf{A}$, the mean of the complex quadratic form $Q_{\mathbf{A}}(\mathbf{Z})$ is given as
\begin{equation}\label{eq: mean Q_A(Z)-1}
\mathrm{E}[ Q_{\mathbf{A}}( \mathbf{Z} ) ] = \mathrm{tr}( \mathbf{A} \boldsymbol{\Sigma}_{\mathbf{Z}} ) + Q_{\mathbf{A}}( \boldsymbol{\mu}_{\mathbf{Z}} )
\end{equation}
where $\boldsymbol{\mu}_{\mathbf{Z}} =\mathrm{E}[\mathbf{Z}]$ and $\boldsymbol{\Sigma}_{\mathbf{Z}} = \mathrm{E}[ \mathbf{Z} \mathbf{Z}^H] - \mathrm{E}[ \mathbf{Z} ] \mathrm{E}[ \mathbf{Y}^H]$
\end{Lem}
\begin{IEEEproof}
The proof is deferred to Appendix~\ref{app: mean complex Q}.
\end{IEEEproof}

Now consider the complex quadratic form $Q_{\mathbf{S}}(\mathbf{Z})$ with a stochastic matrix $\mathbf{S}$ and a complex random vector $\mathbf{Z}$. The following lemma gives the mean of $Q_{\mathbf{S}}(\mathbf{Z})$.
\begin{Lem}\label{lem: mean of Q stochastic A}
For a stochastic and hermitian matrix $\mathbf{S}$, the mean of the quadratic form $Q_{\mathbf{S}}(\mathbf{Z})$ where $\mathbf{Z}$ is a complex random vector with mean $\boldsymbol{\mu}_\mathbf{Z}$ and covariance matrix $\boldsymbol{\Sigma}_\mathbf{Z}$ is given as
\begin{align}\label{eq: E[ Q ]-1}
\mathrm{E}[ Q_{\mathbf{S}}(\mathbf{Z}) ] &= \mathrm{tr}(  \mathrm{E} [\mathbf{S}] ) \boldsymbol{\Sigma}_\mathbf{Z} ) +  Q_{\mathrm{E}[ \mathbf{S}]}( \boldsymbol{\mu_\mathbf{Z}} )
%\boldsymbol{\mu_\mathbf{Z}}^H \mathbb{E}[ \mathbf{S}] \boldsymbol{\mu}_\mathbf{Z} \nonumber \\
%
%\mathit{Var}[ Q_{\mathbf{A}}(\mathbf{X}) ] &= 2 \mathrm{tr}( \boldsymbol{\Sigma}^2 (E_{ \mathbf{A} } [\mathbf{A}] )^2 ) + 4 \boldsymbol{\mu}^T  \boldsymbol{\Sigma} (E_{ \mathbf{A} } [\mathbf{A}] )^2\nonumber.
\end{align}
\end{Lem}

\begin{IEEEproof}
For a given instant of the stochastic matrix $\mathbf{S}$, (\ref{eq: mean Q_A(Z)-1}) provides the conditional mean of the quadratic form $Q_{\mathbf{S}}(\mathbf{Z})$. By taking the expectation over $\mathbf{S}$, we have
\begin{align}\label{eq: mean Q_A(x)}
\mathrm{E}_{\mathbf{S}}[ Q_{\mathbf{S}}(\mathbf{Z}) ] &= \mathrm{E}_{\mathbf{S}} [ \mathrm{tr}( \mathbf{S} \boldsymbol{\Sigma}_\mathbf{Z}) + Q_{\mathbf{S}}(\boldsymbol{\mu}_\mathbf{Z})] \nonumber \\
&= \mathrm{E}_{\mathbf{S}}[ \mathrm{tr}( \mathbf{S} \boldsymbol{\Sigma}_\mathbf{Z}) ] + \boldsymbol{\mu}^H_\mathbf{Z} \mathrm{E} [ \mathbf{S} ] \boldsymbol{\mu}_\mathbf{Z} \nonumber \\
&= \mathrm{tr}( \mathrm{E}[ \mathbf{S} ] \boldsymbol{\Sigma}_\mathbf{Z} ) + \boldsymbol{\mu}_\mathbf{Z}^H \mathrm{E}_{\mathbf{S}} [ \mathbf{S} ] \boldsymbol{\mu}_\mathbf{Z}
\end{align}
where the last line is obtained because the trace is a linear operator. 
\end{IEEEproof}

In using Lemma~\ref{lem: mean of Q stochastic A}, $\boldsymbol{\Pi}_{\mathbf{g}_j}$ needs to be a Hermitian matrix. It is easy to verify that $\boldsymbol{\Pi}_{\mathbf{g}_j}=\boldsymbol{\Pi}_{\mathbf{g}_j}^H$. Moreover, the expectation of $\boldsymbol{\Pi}_{\mathbf{g}_j}$ is required. Denote the random variable $(\mathbf{g}_j^H \mathbf{g}_j)^{-1} \triangleq \Psi$ where the user index $j$ is dropped for brevity. The mean of $\boldsymbol{\Pi}_{\mathbf{g}}$ can be derived as
\begin{align}\label{eq: E[A]}
\mathrm{E}[ \boldsymbol{\Pi}_{\mathbf{g}} ] &= \mathrm{E}[ \mathbf{g} \Psi \mathbf{g}^H ] \nonumber \\
&= \mathrm{E}_{\Psi} [ \mathrm{E}_{\mathbf{g}} [ \mathbf{g} \Psi \mathbf{g}^H ].
%&= \int_0^\infty \mathrm{E}_{\mathbf{g}}[ \psi \mathbf{g} \mathbf{g}^H ] f_{\Psi}(\psi) \mathrm{d} \psi.
\end{align}
Because $\Psi$ is also a function of $\mathbf{g}$, solving (\ref{eq: E[A]}) requires the joint PDF of $\Phi$ and $\mathbf{g}$, which is difficult to obtain.

%====================================
% P_QD here!
%====================================
\begin{figure*}[!t]
% ensure that we have normalsize text
\normalsize
% Store the current equation number. 
\setcounter{MYtempeqncnt}{\value{equation}}
% Set the equation number to one less than the one
% desired for the first equation here.
 %The value here will have to changed if equations
% are added or removed prior to the place these
% equations are referenced in the main text. 
\setcounter{equation}{30}
\begin{align}
\label{eq: QD_integral}
P_{QD} &= \int_0^1 \int_{Q(\theta)}^\infty f_{\Xi}(\xi) \mathrm{d} \xi f_\Theta(\vartheta) \mathrm{d} \vartheta \nonumber \\
&\stackrel{(\mathrm{i})}{=} \int_0^1 \frac{ (1/\theta_S)^{ \alpha_S } (1/\theta_W)^{ \alpha_W } }{ B(\alpha_W, \alpha_S) } \int_{Q(\vartheta)}^\infty \frac{ \xi^{\alpha_W-1} }{ (\frac{1}{\theta_S} + \frac{\xi}{\theta_W} )^{\alpha_W+\alpha_S} }  \mathrm{d} \xi f_\Theta(\vartheta) \mathrm{d} \vartheta \nonumber \\
&\stackrel{(\mathrm{ii})}{=} \int_0^1 \frac{ (Q(\vartheta))^{-\alpha_S} \theta_W^{ \alpha_S } }{ \alpha_S \theta_S^{\alpha_S} B(\alpha_W, \alpha_S) } {}_{2}F_{1}( \alpha_W+\alpha_S, \alpha_S; \alpha_S+1; -\frac{\theta_W}{\theta_S Q(\vartheta)} ) f_\Theta(\vartheta) \mathrm{d} \vartheta \nonumber \\
&\stackrel{(\mathrm{iii})}{=} \left( \frac{ \theta_W }{ \theta_S } \right)^{\alpha_S} \left( \frac{ \theta_W }{ \theta_V } \right)^{\alpha_V} \frac{1}{ \alpha_S B(\alpha_W,\alpha_S) B(\alpha_V,\alpha_W) } \int_0^1 \frac{ {}_{2}F_{1}( \alpha_W+\alpha_S, \alpha_S; \alpha_S+1; -\frac{\theta_W}{\theta_S Q(\vartheta)} ) }{ (Q(\vartheta))^{\alpha_S}\left(1+\frac{ \theta_W }{ \theta_V } \right)^{ \alpha_V+\alpha_W} }  \mathrm{d}\vartheta \nonumber \\
%
%&= \frac{ \beta_S^{\alpha_S} \beta_V^{\alpha_V} }{ \beta_T^{\alpha_S} \alpha_S \beta_Z^{\alpha_V} B(\alpha_T, \alpha_S) B(\alpha_V, \alpha_Z) } \int_0^1 \frac{ _{2}F_{1}( \alpha_T+\alpha_S, \alpha_S; \alpha_S+1; -\frac{\beta_S}{\beta_T Q(z)} ) u^{\alpha_V-1} }{ Q(z)^{\alpha_S}(1+\frac{ \beta_V }{ \beta_Z } )^{ \alpha_V+\alpha_W } } \mathrm{d} u
%P_{QD} = \frac{ \beta_S^{\alpha_S} \beta_V^{\alpha_V} }{ \beta_W^{\alpha_S} \alpha_S \beta_W^{\alpha_V}  B(\alpha_W, \alpha_S) B(\alpha_V, \alpha_W) }%
%\int_0^1  \frac{ {}_2 F_1(\alpha_W+\alpha_S, \alpha_S; \alpha_S+1; -\frac{\beta_S}{\beta_W Q(u)} ) u^{\alpha_V-1}}{ Q(u)^{\alpha_S} (1+\frac{\beta_V}{\beta_W})^{\alpha_V+\alpha_W} } \mathrm{d}u 
\end{align}
% Restore the current equation number. \setcounter{equation}{\value{MYtempeqncnt}}
% The IEEE uses as a separator
\hrulefill
% The spacer can be tweaked to stop underfull vboxes. \vspace*{4pt}
\end{figure*}

\setcounter{equation}{25}

For analytical tractability, we resort to an upper bound of $\mathrm{E}[ \boldsymbol{\Pi}_{\mathbf{g}} ]$ by ignoring the correlation between $\Psi$ and $\mathbf{g}$ such that (\ref{eq: E[A]}) is simplified as
\begin{align}\label{eq: E[A]-2}
\mathrm{E}[ \boldsymbol{\Pi}_{\mathbf{g}} ] %&= E_{\mathbf{g}} [ \mathbf{g} \Psi \mathbf{g}^H ] ] \int_0^\infty \psi f_\Psi (\psi) \mathrm{d} \psi \nonumber \\
= E[ \mathbf{G} ] \cdot E[ \Psi ].
\end{align}
where $\mathbf{G} \triangleq \mathbf{g} \mathbf{g}^H \in \mathbb{C}^{N\times N}$. 
The two expectations in (\ref{eq: E[A]-2}) are derived as follows. %Denote $\mathbf{g}_j \mathbf{g}_j^H \triangleq \mathbf{G}_2=[G_2(m,n)]$, 
According to the structure of $\mathbf{g}$, the $(m,n)$th entry of $\mathbf{G}$ is given by
\begin{align}
 \mathbf{G}&(m,n) = \frac{ \beta K}{K+1} \cos(\varphi_m-\varphi_n)   \nonumber \\
   &+\frac{ \beta \sqrt{K} }{K+1} \Bigl( \cos(\varphi_m) \mathfrak{R}(\mathbf{h}_n) + \cos(\varphi_n) \mathfrak{R}(\mathbf{h}_m) \nonumber \\
&- \sin(\varphi_m) \mathfrak{I}(\mathbf{h}_n) - \sin(\varphi_n) \mathfrak{I}(\mathbf{h}_m) \Bigr) \nonumber \\ 
+ &\frac{\beta}{K+1} \Bigl( \mathfrak{R}(\mathbf{h}_m) \mathfrak{R}(\mathbf{h}_n) + \mathfrak{I}(\mathbf{h}_m) \mathfrak{I}(\mathbf{h}_n) \Bigr) \nonumber \\ 
&+ \Bigl[ \frac{\beta K}{K+1} \sin(\varphi_n - \varphi_m) - \frac{ \sqrt{\beta K} }{ K+1 } \Bigl(\cos(\varphi_m) \mathfrak{I}(\mathbf{h}_n)  \nonumber \\
&+ \sin(\varphi_n) \mathfrak{R}(\mathbf{h}_m) \Bigr)%
 -\frac{\beta}{K+1} \Bigl( \mathfrak{R}(\mathbf{h}_m) \mathfrak{I}(\mathbf{h}_n) \nonumber \\
 & - \mathfrak{I}(\mathbf{h}_m) \mathfrak{R}(\mathbf{h}_n) \Bigr) \Bigr]i
\end{align}
where $\mathbf{h}_n$ denotes the $n$th entry of $\mathbf{h}$. Given that $\mathfrak{R}(\mathbf{h}_m)$, $\mathfrak{I}(\mathbf{h}_m)$, $\mathfrak{R}(\mathbf{h}_n)$, and $\mathfrak{I}(\mathbf{h}_n)$ are independent zero-mean Gaussian random variable with variance $1/2$, $\mathrm{E}[ \mathfrak{R}(\mathbf{h}_m) \mathfrak{R}(\mathbf{h}_n) ]=0$ when $m\neq n$ and $\mathrm{E}[ \mathfrak{R}(\mathbf{h}_m) \mathfrak{R}(\mathbf{h}_n) ]=\frac{1}{2}$ when $m=n$. After some arrangements, the expectation of $\mathbf{G}(m,n)$ is equal to 
\begin{align}\label{eq: mean G2-2}
&\mathrm{E}[\mathbf{G}(m,n)]  \nonumber \\
&= \begin{cases}
\tfrac{\beta K}{K+1} \Bigl( \cos(\varphi_m-\varphi_n) - i\sin(\varphi_m-\varphi_n) \Bigr), 
 &\text{if}~m \neq n\\
 \tfrac{ \beta K}{K+1} \Bigl( \cos(\varphi_m-\varphi_n) -i\sin(\varphi_m-\varphi_n) \Bigr) & \\
 \; + \tfrac{ \beta }{K+1}, &\text{if}~m = n.
\end{cases}
\end{align}

Next, we derive the expectation of $\Psi$. As explained in Sec.~\ref{sec: gamma approx channel power}, $\mathbf{g}^H \mathbf{g}$ can be well approximated to a gamma random variable. This suggests that $\Psi$ follows the inverse gamma distribution (c.f. Lemma~\ref{lem: inverse Gamma dist}). Finally, $\mathrm{E}[\boldsymbol{\Pi}_\mathbf{g}]$ in (\ref{eq: E[A]-2}) is obtained by combining (\ref{eq: Z^-1 statistics}) and (\ref{eq: mean G2-2}).

%====================================
% Variance of Q_Pi(g)
%====================================
\subsection{Variance} %$Q_{\boldsymbol{\Pi}_\mathbf{g}}(\mathbf{g})$
Similar to Sec.~\ref{sec: mean of Q_Pi(g)}, we first derive the variance of the complex quadratic form $Q_{\mathbf{A}}(\mathbf{Z})$ for a deterministic matrix $\mathbf{A}$.
\begin{Lem}\label{lem:  var Q_A(Z)}
Consider a complex random vector $\mathbf{Z}$ where the real and the imaginary parts have the same distribution. For a deterministic matrix $\mathbf{A}$, the variance of the complex quadratic form $Q_{\mathbf{A}}(\mathbf{Z})$ is given as
\begin{equation}\label{eq: var Q_A(Z)}
\mathrm{V}[ Q_{\mathbf{A}}( \mathbf{Z} ) ] = \mathrm{tr}( (\mathbf{A} \boldsymbol{\Sigma}_{\mathbf{Z}})^2 ) + 2Q_{\mathbf{A} \boldsymbol{\Sigma}_\mathbf{Z} \mathbf{A}}( \boldsymbol{\mu}_{\mathbf{Z}}^H )
\end{equation}
where $\boldsymbol{\mu}_{\mathbf{Z}} =\mathrm{E}[\mathbf{Z}]$ and $\boldsymbol{\Sigma}_{\mathbf{Z}} = \mathrm{E}[ \mathbf{Z} \mathbf{Z}^H] - \mathrm{E}[ \mathbf{Z} ] \mathbb{E}[ \mathbf{Y}^H]$
\end{Lem}
\begin{IEEEproof}
The proof is deferred to Appendix~\ref{app: lem:  var Q_A(Z)}.
\end{IEEEproof}
The case when $\mathbf{A}$ is a stochastic matrix is addressed below.
\begin{Lem}\label{lem:var complex stochastic Q}
For a stochastic and hermitian matrix $\mathbf{S}$, the variance of the quadratic form $Q_{\mathbf{S}}(\mathbf{Z})$ where $\mathbf{Z}$ is a complex random vector with mean $\boldsymbol{\mu}_\mathbf{Z}$ and covariance matrix $\boldsymbol{\Sigma}_\mathbf{Z}$ is given as
\begin{equation}\label{eq: var Q_A(Z)}
\mathrm{V}[ Q_{\mathbf{S}}(\mathbf{Z}) ] = \mathrm{tr}\bigl( \boldsymbol{\Sigma}_{\mathbf{Z}}^2 ( \mathrm{E} [\mathbf{S}] )^2 \bigr) + 2 Q_{ \boldsymbol{\Sigma}_{\mathbf{Z}} (\mathrm{E} [\mathbf{S}] )^2 }(\boldsymbol{\mu}_{\mathbf{Z}}).
\end{equation}
\end{Lem}
\begin{IEEEproof}
The proof is deferred to Appendix~\ref{app:var complex stochastic Q}.
\end{IEEEproof}

\begin{spacing}{0.6}
\begin{algorithm}
\footnotesize
\caption{Approximated QD probability} \label{alg:qd}
\algnewcommand\algorithmicto{\textbf{to}}
%\algrenewtext{For}[3]%
%{\algorithmicfor\ #1 $\gets$ #2 \algorithmicto\ #3 \algorithmicdo}
  \begin{algorithmic}[2]
  \Require 
     \State User channels $\mathbf{g}_i$ and $\mathbf{g}_j$; number of antennas $N$.
%------ 
 \For{$l \gets i, j$}
 \For{$n \gets 1$ to $N$}
  \State Approxiate $\mathfrak{R}(\mathbf{g}_l[n])$ as $\Gamma(k_{\text{real},\mathbf{g}_l[n] }, \theta_{\text{real},\mathbf{g}_l[n]})$;
  \State Approximate $\mathfrak{I}(\mathbf{g}_l[n])$ as $\Gamma(k_{\text{imag},\mathbf{g}_l[n]}, \theta_{\text{imag},\mathbf{g}_l[n]})$;
  \EndFor  
  \EndFor
  %---
   \State Approximate $\norm{\mathbf{g}_i}^2$ as $W\sim \Gamma(\alpha_W, \theta_W)$ using (\ref{eq: order matching parameters});
   %---
   \State Approximate $\norm{\mathbf{g}_j}^2$ as $S\sim \Gamma(\alpha_S, \theta_S)$ using (\ref{eq: order matching parameters});
    \State Approximate $Q_{ \boldsymbol{\Pi}_{\mathbf{g}_j} }( \mathbf{g}_i )$ as $V\sim \Gamma(k_V, \theta_V)$ according to Sec.~\ref{sec: gamma-quadratic-form};
  \State Compute $P_{QD}$ using (\ref{eq: QD final});
  \end{algorithmic}
\end{algorithm}
\end{spacing}

\section{Approximated Quasi-Degradation Probability}\label{sec: approx-qd-prob}
With the PDFs for $\Xi$ and $\Theta$ obtained through the gamma approximation, the QD probability in (\ref{eq:P_QD_orig}) can be derived as (\ref{eq: QD_integral}) on the top of this page, where (i) is obtained by using (\ref{eq: Beta prime}); (ii) is reached with the help of~\cite[(3.194-2)]{Gradshteyn2007} where $ _{2}F_{1}(a, b; c; z)$ denotes the Gauss hypergeometric function. Finally, (iii) is obtained by approximating $\Theta$ as the ratio of two gamma random variables with the PDF given in  (\ref{eq: Beta prime}). The definite integral in (\ref{eq: QD_integral}) can be simplified using the series representation of $ _{2}F_{1}(a, b; c; z)$~\cite[9.10]{Gradshteyn2007}, leading to

\addtocounter{equation}{1}

\begin{align}\label{eq: QD final}
 P_{QD}^{(i,j)} =& \sum_{k=0}^\infty \frac{ (\alpha_W+\alpha_S)_k (\alpha_S)_k }{ (\alpha_S+1)_k k! } (-\frac{ \theta_W }{ \theta_S })^k \nonumber \\
 &\times \left(1+\frac{\theta_W}{\theta_V} \right)^ {-(\alpha_V+\alpha_W)}
 G(k, r_i, r_j, \alpha_S, \vartheta)
\end{align}
where $(a)_k = a(a+1)\cdots(a+k)$ and $G(k, r_i, r_j, \alpha_S, \vartheta) = \int_0^1 \left( \tfrac{1+r_i}{\vartheta}-\frac{r_i \vartheta}{(1+r_j(1-\vartheta))^2} \right)^{-(k+\alpha_S)} \mathrm{d} \vartheta$. For readers' convenience, the procedure for computing the QD probability is summarized in Algorithm~\ref{alg:qd}. 

As to the case of more than two users,  the exact analysis for the QD probability is subject to future work. A conservative lower-bound in the pairwise sense can be found as
\begin{equation}
P_{QD} \geq \sum_{i=1}^K \sum_{j=1}^{K-1} P_{QD}^{(i,j)} 
\end{equation}
where $K$ is the number of users and $P_{QD}^{(i,j)}$ denotes the QD probability between the $i$th and the $j$th user.

\section{Numerical Results}\label{sec:results}

Numerical results are presented in this section to evaluate the QD probability subject to different system parameters. The accuracy of the proposed analysis for the QD probability is also validated. Without loss of generality, consider two arbitrary users and they are assigned with the index $i=1$ and $j=2$. Table~\ref{table: simulation-parameters} lists the simulation parameters. To reflect the difference of the channel strength, define the path-loss ratio of the two user channels as $\beta_\Delta=\beta_1/\beta_2$. A larger value of $\beta_\Delta$ mimics the scenario that the two NOMA users have very different channel gains. By choosing $\beta_\Delta \geq 1$, we can ensure that decoding user 1's signal first satisfies the necessary condition of QD. %Finally, the target rate constraint is set as $r_1=r_2=1$. 

%=== Fig. 2 ===
\begin{figure}[!t]
\centering
\includegraphics[width=1.0\columnwidth]{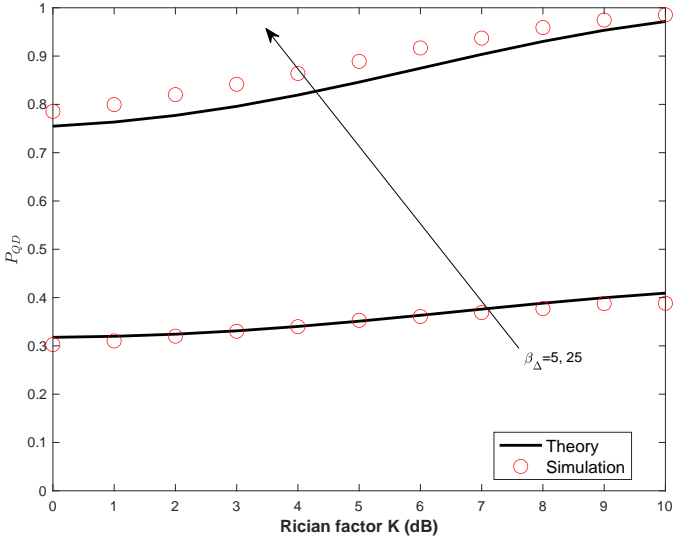}
\caption{QD probability vs. Rician factor $K$ for varied path-loss ratio $\beta_\Delta$ and fixed angle difference $\theta_\Delta=10$.} \label{fig:Pqd vs B}
\end{figure}

Fig.~\ref{fig:Pqd vs B} plots the QD probability versus Rician factor $K$ with $\beta_\Delta=5$ and 25, respectively. One can see that when the user channels are more LOS dominant (i.e. larger $K$), the QD probability is higher and the increasing trend is more remarkable if the two user channels are more different in their strengths (i.e., larger $\beta_\Delta$). For example, the QD probability with $\beta_\Delta=5$ is about half of that with $\beta_\Delta=25$. This agrees with the known results that NOMA gain is more pronounced when the two NOMA users have more different channel strengths. Notice that the above discussions are obtained with a fixed angular difference between two user channels. The increasing trend shown in Fig.~\ref{fig:Pqd vs B} does not always hold, which will be illuminated later. In terms of the analysis accuracy, the analytical results mostly match to the simulated ones. The discrepancy revealed on the figure is the consequence of the approximated distributions used in the analysis. Since various approximations are employed, their impacts to the analysis accuracy will be examined in the subsequent discussions.

\begin{table}[!t]
% increase table row spacing, adjust to taste
%\renewcommand{\arraystretch}{1}
% if using array.sty, it might be a good idea to tweak the value of
% \extrarowheight as needed to properly center the text within the cells
\caption{Default simulation parameters.}
\label{table: simulation-parameters}
\centering
% Some packages, such as MDW tools, offer better commands for making tables
% than the plain LaTeX2e tabular which is used here.
\resizebox{0.8\columnwidth}{!}{%
\begin{tabular}{|c||c|}
\hline
Azimuth angle & $\theta_1=30^\circ$, $\theta_2 = \theta_1+\theta_\Delta$ \\
\hline
Number of BS antennas & $N=4$ \\
\hline
Pathloss  ratio & $\beta_\Delta=\beta_1/\beta_2$\\
\hline
Rician factor & $K=0\sim 10$ dB~\cite{Samim2016} \\
\hline
Target rate & $r_1=r_2=1$~\cite{Chen2016} \\
\hline
\end{tabular}
}
\end{table}

%=== Fig. 3 ===
\begin{figure}[!t]
\centering
\includegraphics[width=1.0\columnwidth]{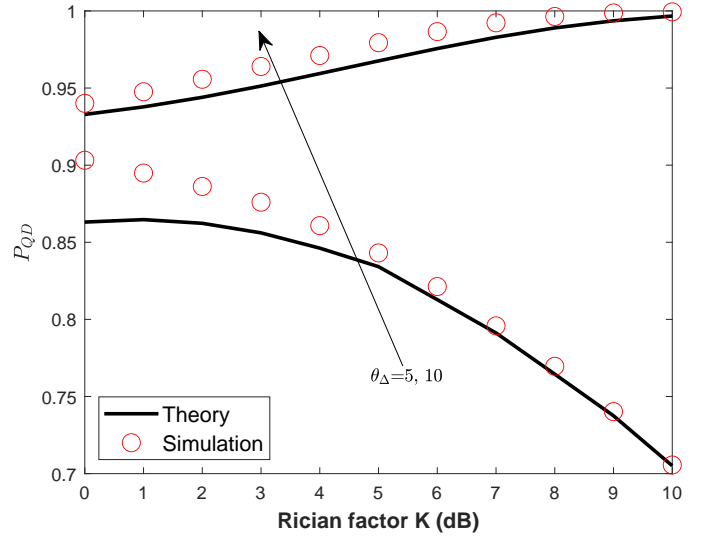}
\caption{QD probability vs. Rician factor $K$ for varied angle difference $\theta_\Delta$ and fixed pathloss ratio $\beta_\Delta=100$.} \label{fig:Pqd vs theta}
\end{figure}

The impact of the uesr's angular difference to the QD probability is investigated in Fig.~\ref{fig:Pqd vs theta} where the QD probability is plotted as a function of Rician factor $K$ for $\theta_\Delta=5$ and 10. Here, $\beta_\Delta$ is fixed to 100. A small angular difference implies that the two users are close in the angular domain and thus they are likely to be served by the same transmitting beam using the typical beam selection algorithm. It is interesting to see that when the angular difference is small, i.e., $\theta_\Delta=5$, the QD probability decreases with $K$, which is opposed to the case when $\theta_\Delta=10$. This suggests that when the two users are close in their azimuth angles and their channels are LOS-dominated (namely, $K$ is large), the probability for their channels to be quasi-degraded becomes small. This is true even the two user channels are very different in strength (e.g, $\beta_\Delta=100$). Consequently, NOMA is not preferable because the chance for NOMA to achieve the same performance as DPC is diminished. On the other hand, NOMA can be beneficial to serve the users with close azimuth angles if the LOS strengths of their channels are not significant (e.g., $\theta_\Delta=5$ and $K$ is small), yet the QD probability remains lower than the case with a larger angular difference ($\theta_\Delta=10$). Here, we observe a reasonable match between the analytical results and the simulated ones except when $K$ and $\theta_\Delta$ are small. The cause will be discussed next.

%=== Fig. 4 ===
\begin{figure}[!t]
\centering
\includegraphics[width=1.0\columnwidth]{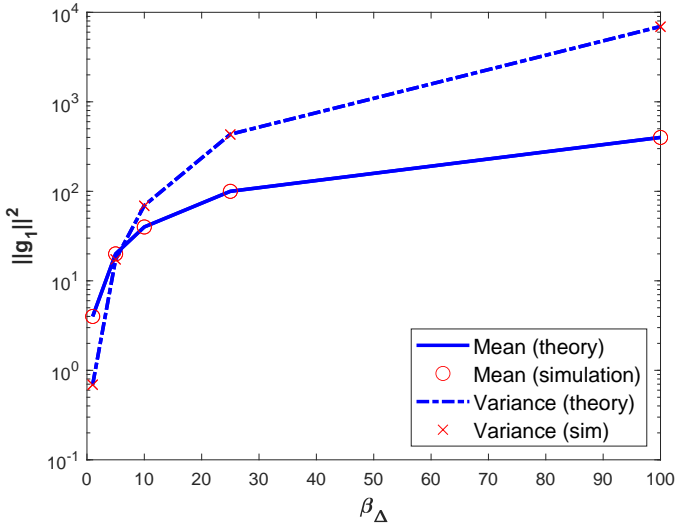}
\caption{Mean and variance of $\norm{\mathbf{g}_1}^2$ for $K=10$ dB.} \label{fig:meang1norm}
\end{figure}

Since the exact analysis for the QD probability is not tractable, several approximations are employed in this work. We first examine the gamma approximation for the channel powers because this is the root that leads us to a tractable analysis for the QD probability. Fig.~\ref{fig:meang1norm} plots the theoretical mean and variance obtained by first computing the shape and scale parameters of a gamma random variable used to approximate $\norm{\mathbf{g}_1}^2$ according to Remark~\ref{lem: order matching}. Then Lemma~\ref{lem: gamma distri} is used to compute the required theoretical mean and variance. 
%used to approximate the numerator and the denominator of $\Theta$ defined in (\ref{eq: revised u}), both involving the the Gamma approximation for $\norm{\mathbf{g}_1}^2$ using Lemma~\ref{lem: order matching}. 
The analytical results are compared with the simulated ones for $\beta_\Delta \in [1,100]$ and $K=10$ dB. As shown, the gamma approximation for $\norm{\mathbf{g}_1}^2$ is promising to capture the first two moments. Both the mean and the variance of $\norm{\mathbf{g}_1}^2$ increase with $\beta_\Delta$. The increasing trend can be explained by observing the mean and the covariance matrix of the channel vector $\mathbf{g}$ given by
\begin{align}\label{eq: mean var Rician}
\boldsymbol{\mu}_{\mathbf{g}_1} &= \sqrt{ \frac{\beta K}{K+1} }\mathbf{a} \\
\boldsymbol{\Sigma}_{\mathbf{g}_1} &= \frac{\beta}{K+1} \mathbf{I}_N.
\end{align}
Since each entry in $\boldsymbol{\mu}_{\mathbf{g}_1}$ is proportional to $\sqrt{\beta}$, the mean of $\norm{\mathbf{g}_1}^2$ is an increasing function of $\beta$. Likewise, $\boldsymbol{\Sigma}_{\mathbf{g}_1}$ increases with $\beta$ and so does the variance of $\norm{\mathbf{g}_1}^2$.

%=== Fig. 5 ===
\begin{figure}[!t]
\centering
\includegraphics[width=1.0\columnwidth]{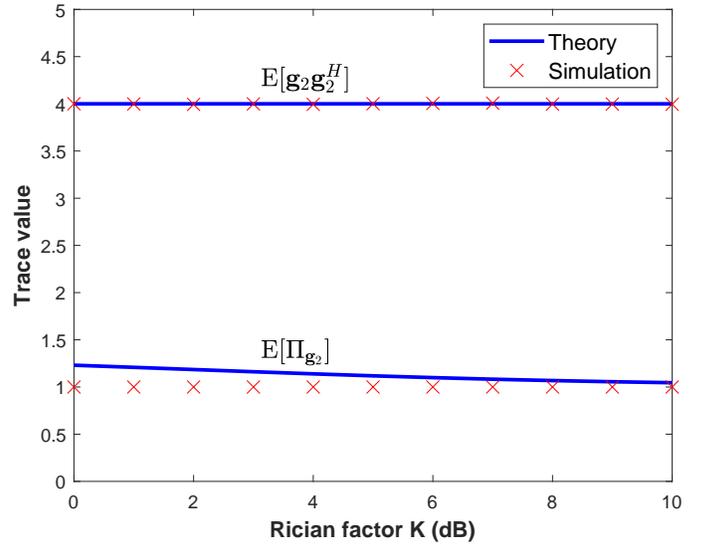}
\caption{Trace values of two square matrices: $\mathrm{E}[\mathbf{g}_2 \mathbf{g}_2^H]$ and $\mathrm{E}[\boldsymbol{\Pi}_{\mathbf{g}_2}]$.} \label{fig: trace}
\end{figure}

Next, we assess the approximation for the numerator of $\Theta$, which is the angle between two user channels. As explained in the beginning of Sec.~\ref{sec: gamma-quadratic-form}, the numerator of $\Theta$ appears in a matrix quadratic form $Q_{\boldsymbol{\Pi}_{\mathbf{g}_2}}(\mathbf{g}_1)$ and can be approximated as a gamma random variable with the mean the variance given in  (\ref{eq: E[A]-2}) and (\ref{eq: var Q_A(Z)}), respectively. The mean in (\ref{eq: E[A]-2}) is is approximately equal to the product of $\mathrm{E}[\mathbf{g}_2 \mathbf{g}_2^H]$ and $\mathrm{E}[\Psi]$ by ignoring the dependence between $\Psi$ and $\mathbf{g}_2$. For $\Psi$, it is the inverse of $\norm{\mathbf{g}_2}^2$ whose mean and variance closely follow those of the gamma distribution as revealed in Fig.~\ref{fig:meang1norm}. To validate the accuracy of  (\ref{eq: E[A]-2}),  Fig.~\ref{fig: trace} plots the trace values for $\mathrm{E}[\mathbf{g}_2 \mathbf{g}_2^H]$ and $\mathrm{E}[\boldsymbol{\Pi}_{\mathbf{g}_2}]$, both being a square matrix, for $K\in[0, 10]$ dB. It can be seen that the theoretical trace values of $\mathrm{E}[\mathbf{g}_2 \mathbf{g}_2^H]$ using (\ref{eq: mean G2-2}) perfectly match with the simulated ones while the theoretical trace values of $\mathrm{E}[\boldsymbol{\Pi}_{\mathbf{g}_2}]$ using (\ref{eq: E[A]-2}) slightly deviate from the simulated ones when $K$ is small. With a smaller Rician factor $K$, the user channel is more sensitive to the dynamics in the NLOS component and thus ignoring the dependence between $\Psi$ and $\mathbf{g}_2$ introduces errors in evaluating the mean of $\boldsymbol{\Pi}_{\mathbf{g}_2}$.

%=== Fig. 6 ===
\begin{figure}[!t]
\centering
\includegraphics[width=1.0\columnwidth]{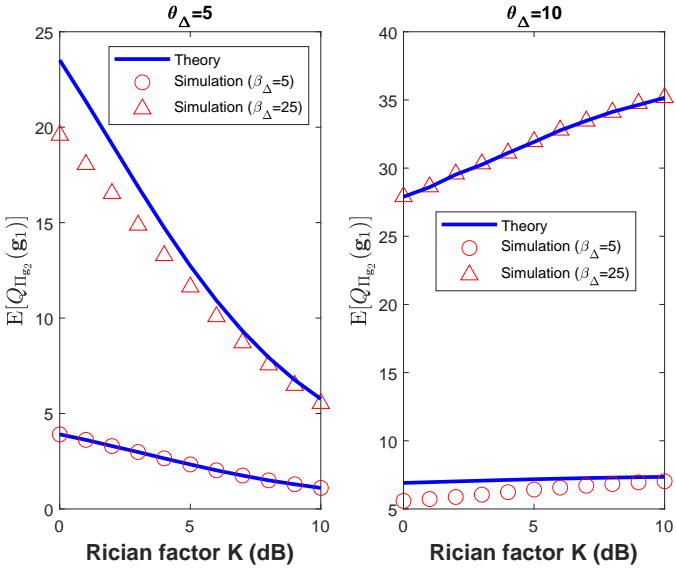}
\caption{Means of $Q_{\boldsymbol{\Pi}_{\mathbf{g}_2}}(\mathbf{g}_1)$. Left: $\theta_\Delta=5$; right: $\theta_\Delta=10$.} \label{fig:meanZ}
\end{figure}

Finally, we validate the accuracy of the approximated mean and variance of the complex quadratic form $Q_{\boldsymbol{\Pi}_{\mathbf{g}_2}}(\mathbf{g}_1)$. In Fig.~\ref{fig:meanZ}, the theoretical and simulated mean values are plotted as a function of $K$ for varied $\theta_\Delta$ and $\beta_\Delta$. One can see that the analytical mean values match to the simulated ones, except the case with a smaller angle difference $\theta_\Delta$ and the Rician factor $K$. Under both conditions, (\ref{eq: E[A]-2}) is more loose in approximating the true expectation in (\ref{eq: E[A]}\hl) due to the ignored dependence between $\Psi$ and $\mathbf{g}$ as explained above. It is also noticed that the curve of $\mathrm{E}[Q_{\boldsymbol{\Pi}_{\mathbf{g}_2}}(\mathbf{g}_1)]$ in Fig.~\ref{fig:meanZ} follows the same trend as the QD probability in Fig.~\ref{fig:Pqd vs theta}. Meanwhile, $Q_{\boldsymbol{\Pi}_{\mathbf{g}_2}}(\mathbf{g}_1)$ is proportional to $\Theta$, according to (\ref{eq: revised u}). Consequently, the QD probability is proportional to $\Theta$, the angle between channel vectors. 

\begin{figure}[!t]
\centering
\includegraphics[width=1.0\columnwidth]{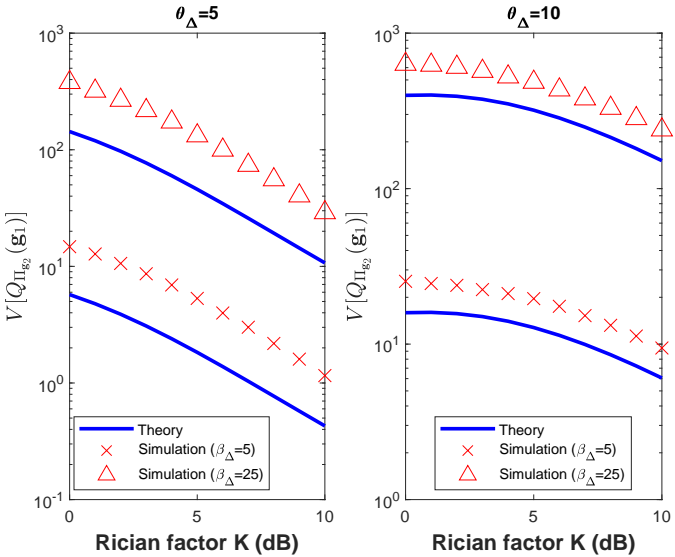}
\caption{Variances of $Q_{\boldsymbol{\Pi}_{\mathbf{g}_2}}(\mathbf{g}_1)$. Left: $\theta_\Delta=5$; right: $\theta_\Delta=10$.} \label{fig:varZ}
\end{figure}

The variance of $Q_{\boldsymbol{\Pi}_{\mathbf{g}_2}}(\mathbf{g}_1)$ is plotted in Fig.~\ref{fig:varZ}. Comparing with the mean values shown in Fig.~\ref{fig:meanZ}, a larger difference between the theoretical variances and the simulated ones is revealed. This is because the variance computed from (\ref{eq: var Q_A(Z)}) involves the square of $\mathrm{E}[\boldsymbol{\Pi}_{\mathbf{g}_2}]$ that pronounces the approximation error. Regardless the angle difference and the pathloss ratio, the variance decreases with $K$. This is because the variance is mainly caused by the NLOS component and thus it becomes smaller when the channel is more LOS dominant (i.e., larger $K$).

%=== Fig. 7 ===

\section{Conclusion}\label{sec:summary}
For MISO Rician fading channels, an analytical framework is proposed to derive the QD probability that characterizes the optimality of NOMA in approaching the capacity region of the two-user broadcast channel. 
The QD probability of interest involves a matrix quadratic form whose exact distribution is not available. With the aid of a series of approximations based on the gamma distribution, we obtained the QD probability over MISO Rician fading channels in closed form. Our work is versatile in capturing important channel parameters including both the large-scale and the small-scale fading, the array factors, and the angular information of LOS paths. Numerical results indicate that the obtained expression is accuracy for a wide range of the Rician factor and the angle difference between two users. Our results also reveal the coupled impact of channel angles and LOS dominance. 
\begin{itemize}
\item Unlike the Rayleigh fading channels that permit a high QD probability as long as two users have very different channel gains, the QD probability in the Rician fading channels is diminished if the two user channels are LOS dominant and close in the angular domain. This is true even their channel gain difference is large.
\item When two users are close in the angular domain, NOMA is possible to achieve the same performance as DPC with a high probability ($>85\%$) if their LOS paths are not dominant ($K<5$ dB) and the channel gain difference is sufficiently large.
\item The QD probability is proportional to the angle between two channel vectors and this result holds true for both LOS dominant and non-dominant fading channels.
\end{itemize}
The results of our work may find some useful applications. For example, user grouping is essential to NOMA systems and the QD probability can be used to assess if the channels of potential NOMA users are likely to be quasi-degraded. In the emerging aerial-ground communications, the cellular-connected UAVs may access the cellular band using NOMA. Since UAV-to-BS channels are mostly likely dominated by LOS, the BS should use NOMA to serve two or more UAVs carefully without degrading the QD probability. Also, the matrix quadratic form commonly appears in the performance metric of multi-antenna wireless systems. One can approximate the matrix quadratic form in random vectors with non-central distributions to the gamma distribution with acceptance accuracy. Another interesting direction is to replace the gamma approximation used in this work by the Nakagami random variable, based on the fact that the squared Nakagami random variable is the gamma random variable. While the Nakagami model is widely used to capture different fading conditions, the channel power gain of the MISO channel becomes the sum of squared non-identically distributed Nakagami random variables. Thus the analysis for QD probability using the Nakagami model is not trivial and deserves further work.

\renewcommand{\theequation}{A.\arabic{equation}}
\appendices%
\setcounter{equation}{0}

%=====================================================================
% Appendix A
%=====================================================================
\section{Proof for Lemma~\ref{lem: X^2 distribution}}\label{app: squared gaussian approx}
Denote $Y=(\frac{X}{\sigma})^2$. For $X \sim \mathcal{N}(\mu, \sigma^2)$, it follows that %$(\frac{G-\mu}{\sigma})^2$ is Chi-squared distributed with one degree of freedom, i.e., $(\frac{G-\mu}{\sigma})^2 \sim \chi^2(1)$. For convenience, denote 
$Y$ is non-central $\chi^2$ distributed with one degree of freedom and the non-centrality parameter equal to $ (\frac{\mu}{\sigma})^2$. Consequently, the mean and variance of $Y$ are given by $(1+(\frac{\mu}{\sigma})^2)$ and $2(1+2(\frac{\mu}{\sigma})^2)$, respectively. Since $X^2 = \sigma^2 Y$, the mean and variance of $X^2$ can be found as
\begin{align}
\mathrm{E}[X^2] &=  \mathrm{E}[ \sigma^2 Y] = \sigma^2+\mu^2\\
\mathrm{V}[X^2] &= \mathrm{V}[ \sigma^2 Y] = 2\sigma^2(\sigma^2+2\mu^2).
\end{align} 
Using the first two moments and Lemma~\ref{lem: gamma distri}, (\ref{eq: X^2 gamma}) can be obtained that completes the proof.

%=====================================================================
% Appendix B
%=====================================================================
\section{Proof for Lemma~\ref{lem: mean Q_A(Z)}}\label{app: mean complex Q}
Following Lemma~\ref{def: complex rv}, the complex random vector $\mathbf{Z}$ can be constructed from a pair of real random vectors $\mathsf{X}=(\mathbf{X}_1^T,\mathbf{X}_2^T)^T$.  When $\mathbf{X}_1$ and $\mathbf{X}_2$ are drawn from the same distribution, they have the same means and covariance matrices, i.e., 
\begin{align}\label{eq: identical statistics}
\mathrm{E}[\mathbf{X}_1] &= \mathbb{E}[\mathbf{X}_2]=\boldsymbol{\mu} \nonumber \\
\boldsymbol{\Sigma}_{\mathbf{X}_1} &= \boldsymbol{\Sigma}_{\mathbf{X}_2} = \boldsymbol{\Sigma}.
\end{align}
%By extending the result in Lemma~\ref{lem: statistics of Q} for the real random vector to the complex random vector, the mean of the complex quadratic form $Q_{\mathbf{A}}( \mathbf{Z} )$ can be obtained as (\ref{eq: mean Q_A(Z)-1}).
%
Since $\mathbf{A}$ is symmetric, it can be shown that the complex quadratic form $Q_{\mathbf{A}}(\mathbf{Z})$ is connected through the real quadratic form through the following equation.
\begin{align}
Q_{\mathbf{A}}(\mathbf{Z}) &= (\mathbf{X}_1^T - j \mathbf{X}_2^T) \mathbf{A} (\mathbf{X}_1 + j \mathbf{X}_2) \nonumber \\
&= Q_{\mathbf{A}}( \mathbf{X}_1 ) + Q_{\mathbf{A}}( \mathbf{X}_2 )
\end{align}
Therefore, 
\begin{align}\label{eq: mean Q_A(Z)-2}
\mathrm{E}[ Q_{\mathbf{A}}(\mathbf{Z}) ] &= \mathrm{E}[Q_{\mathbf{A}}(\mathbf{X}_1)] +   \mathrm{E}[Q_{\mathbf{A}}(\mathbf{X}_2)]\nonumber \\
&= 2\mathrm{tr}( \mathbf{A} \boldsymbol{\Sigma} ) + 2 Q_{\mathbf{A}}( \boldsymbol{\mu} ).
%%\mathbb{V}[ Q_{\mathbf{A}}(\mathbf{Z}) ] &= 4 \mathbb{V}[Q_{\mathbf{A}}(\mathbf{X})],
\end{align}
which is obtained using Lemma~\ref{lem: statistics of Q} and the notations defined in (\ref{eq: identical statistics}). Let's work on the covariance matrix of $\mathbf{Z}$, denoted by $\boldsymbol{\Sigma}_{\mathbf{Z}}$. Using (\ref{eq: complex pair}), we have
\begin{align}\label{eq: Sigma_Z}
\boldsymbol{\Sigma}_{\mathbf{Z}} &= \mathrm{E}[ \mathbf{Z} \mathbf{Z}^H ] - \mathrm{E}[\mathbf{Z}]\mathrm{E}[\mathbf{Z}^H] \nonumber \\
&= 4 \mathbf{M}^H \boldsymbol{\Sigma} \mathbf{M}.
\end{align}
Therefore, 
\begin{align}\label{eq: tr Sigma_Z}
\mathrm{tr}( \boldsymbol{\Sigma}_{\mathbf{Z}} ) &= 4\mathrm{tr}( \mathbf{M}^H \boldsymbol{\Sigma} \mathbf{M} ) \nonumber \\
&= 4 \mathrm{tr}( \boldsymbol{\Sigma} \mathbf{M}^H \mathbf{M}) \nonumber \\
&= 2 \mathrm{tr}( \boldsymbol{\Sigma})
\end{align}
where we have used the fact that $\mathbf{M}^H \mathbf{M}=\frac{1}{2} \mathbf{I}_N$. By multiplying $\boldsymbol{\Sigma}_{\mathbf{Z}}$ with matrix $\mathbf{A}$, we have
\begin{align}\label{eq: tr A Sigma_Z}
\mathrm{tr}( \mathbf{A} \boldsymbol{\Sigma}_\mathbf{Z} ) = 2 \mathrm{tr}( \mathbf{A} \boldsymbol{\Sigma} ).
\end{align}
On the other hand, we can establish the quadratic form $Q_{\mathbf{A}}( \boldsymbol{\mu}_{\mathbf{Z}})$ in terms of the real-numbered vectors $\boldsymbol{\mu}$ as
\begin{align}\label{eq: Q_A(mu_Z)}
Q_{\mathbf{A}}( \boldsymbol{\mu}_{\mathbf{Z}}) &= (\boldsymbol{\mu}^T-j \boldsymbol{\mu}^T ) \mathbf{A}  (\boldsymbol{\mu} + j \boldsymbol{\mu} ) \nonumber \\
&= 2  \boldsymbol{\mu}^T \mathbf{A} \boldsymbol{\mu} \nonumber \\
&= 2 Q_{\mathbf{A}}( \boldsymbol{\mu} ).
\end{align}
Based on (\ref{eq: tr A Sigma_Z}) and (\ref{eq: Q_A(mu_Z)}), (\ref{eq: mean Q_A(Z)-2}) can be rewritten as (\ref{eq: mean Q_A(Z)-1}) and the proof is completed.

%=====================================================================
% Appendix C
%=====================================================================
\section{Proof for Lemma \ref{lem:  var Q_A(Z)}}\label{app: lem:  var Q_A(Z)}
Using the same argument for obtaining  (\ref{eq: mean Q_A(Z)-2}), the variance of $Q_{\mathbf{A}}(\mathbf{Z})$ can be given as
\begin{align}\label{eq: var-Q_A(Z)-1}
\mathrm{V}[ Q_{\mathbf{A}}(\mathbf{Z}) ] &= \mathrm{V}[ Q_{\mathbf{A}}( \mathbf{X}_1 ) ] + \mathrm{V}[ Q_{\mathbf{A}}( \mathbf{X}_2 ) ] \nonumber\\
&=  4 \mathrm{tr}( (\mathbf{A} \boldsymbol{\Sigma} )^2 ) + 8 Q_{ \mathbf{A} \mathbf{\Sigma} \mathbf{A} }(\boldsymbol{\mu} ).
\end{align}
With the aid of (\ref{eq: tr A Sigma_Z}), it can be established that
\begin{equation}\label{eq: tr (A Sigma)^2}
\mathrm{tr}( (\mathbf{A} \boldsymbol{\Sigma}_{\mathbf{Z}} )^2 ) %
%= \mathrm{tr}( \mathbf{A} \boldsymbol{\Sigma}_{\mathbf{Z}} \mathbf{A}^H \boldsymbol{\Sigma}_{\mathbf{Z}}  ) %\nonumber \\
= 4\mathrm{tr}( (\mathbf{A} \boldsymbol{\Sigma} )^2 ).
\end{equation}
In addition, the complex quadratic form $Q_{ \mathbf{A} \boldsymbol{\Sigma}_\mathbf{Z} \mathbf{A} }(\boldsymbol{\mu}_\mathbf{Z}^H)$ can be expressed in terms of the real quadratic form as
\begin{align}\label{eq: Q_A Sigma A (mu)}
Q_{ \mathbf{A} \boldsymbol{\Sigma}_\mathbf{Z} \mathbf{A} } (\boldsymbol{\mu}_\mathbf{Z}^H) &=%
(\boldsymbol{\mu} + j\boldsymbol{\mu})  \mathbf{A} (4 \mathbf{M}^H \boldsymbol{\Sigma}\mathbf{M} )( \boldsymbol{\mu}^T - j \boldsymbol{\mu}^T ) \nonumber \\
&= 4\boldsymbol{\mu} \mathbf{A} \boldsymbol{\Sigma} \mathbf{A} \boldsymbol{\mu}^T \nonumber \\
&= 4Q_{\mathbf{A} \boldsymbol{\Sigma} \mathbf{A} } (\boldsymbol{\mu}^T),
\end{align}
which is obtained because $\mathbf{M}^H \mathbf{M} = \frac{1}{2} \mathbf{I}_N$. Using (\ref{eq: tr (A Sigma)^2}) and (\ref{eq: Q_A Sigma A (mu)}), (\ref{eq: var-Q_A(Z)-1}) can be rewritten as (\ref{eq: var Q_A(Z)}) that completes the proof.

%=====================================================================
% Appendix D
%=====================================================================
\section{Proof for Lemma~\ref{lem:var complex stochastic Q}}\label{app:var complex stochastic Q}
The variance of $Q_{\mathbf{S}}(\mathbf{Z})$ for stochastic $\mathbf{S}$ can be found as
\begin{align}
\mathrm{V}[ Q_{\mathbf{S}}(\mathbf{Z}) ] &= \mathrm{E}_{ \mathbf{S} }\bigl[ \mathbb{V}[ Q_{\mathbf{S}}(\mathbf{Z}) ] \bigr] \nonumber \\
&=  \mathrm{E}_{ \mathbf{S} }\bigl[ \mathrm{tr}( (\mathbf{S} \boldsymbol{\Sigma}_{\boldsymbol{Z}} )^2 ) + 2 Q_{\mathbf{S} \boldsymbol{\Sigma}_{\mathbf{Z}} \mathbf{S} }( \boldsymbol{\mu}_{\mathbf{Z}}) \bigr] \nonumber \\
&\stackrel{(\mathrm{i})}{=} \mathrm{tr}\bigl( \mathrm{E}_{ \mathbf{S} }[ (\mathbf{S} \boldsymbol{\Sigma}_{\mathbf{Z}} )^2 ] \bigr) + 2\boldsymbol{\mu}_{\mathbf{Z}}^H \mathrm{E}_{ \mathbf{S} } [ \mathbf{S} \boldsymbol{\Sigma}_{\mathbf{Z}} \mathbf{S} ] \boldsymbol{\mu}_{\mathbf{Z}} \nonumber \\
&\stackrel{\mathrm{(ii)}}{=} \mathrm{tr}\bigl( \boldsymbol{\Sigma}_{\mathbf{Z}}^2 ( \mathrm{E} [\mathbf{S}] )^2 \bigr) + 2 \boldsymbol{\mu}_{\mathbf{Z}}^H  \boldsymbol{\Sigma}_{\mathbf{Z}} (\mathrm{E} [\mathbf{S}] )^2 \boldsymbol{\mu}.
\end{align}
where (i) is obtained by exchanging the expectation and the trace operations due to the linearity of the trace operator and (ii) is obtained because the covariance $\boldsymbol{\Sigma}_{\mathbf{Z}}$ is constant to the expectation over $\mathbf{A}$. Finally, (\ref{eq: var Q_A(Z)}) is obtained by writing the last term in the quadratic form.

\bibliographystyle{IEEEtran}

\begin{IEEEbiography}{Kuang-Hao (Stanley) Liu} (S'06--M'08) 
received the Ph.D. degree in electrical and computer engineering from the University of Waterloo, Canada, in 2008. From 2000 to 2002, he was an Engineer with Siemens Telecom System Ltd., Taiwan. From 2004 to 2008, he was a Research Assistant with the Broadband Communications Research Group, University of Waterloo. He is currently a Professor with the Department of Electrical Engineering, National Cheng Kung University, Tainan, Taiwan. His recent research focuses on cooperative communications, wireless energy-harvesting, and mmWave communications.

Dr. Liu has served as a technical program committee member for many IEEE conferences, such as the IEEE International Conference on Communications and the IEEE Global Telecommunications Conference. He was a recipient of the Best Paper Award from the IEEE Wireless Communications and Networking Conference 2010. He has participated in organizing several international conferences, including the Chinacom 2009 and 2010, the Wicon 2010, the IEEE PIMRC 2012, and the IEEE SmartGridComm 2012. He was a Guest Editor of the IET Communications Special Issue on Secure Physical Layer Communications and an Editor of the IEEE Wireless Communications Magazine.
\end{IEEEbiography}

\end{document}